\documentclass[preprint,nofootinbib,showkeys,showpacs,tightenlines,fleqn,prd]{revtex4}        

\usepackage{graphicx} 
\begin{document}      
\preprint{PNU-NTG-05/2006} 
\preprint{PNU-NURI-06/2006} 
\title{Leading-twist pion and kaon distribution amplitudes   \\ 
in the gauge-invariant nonlocal chiral quark model \\ 
from the instanton vacuum}  
\author{Seung-il Nam} 
\email{sinam@pusan.ac.kr} 
\affiliation{Department of 
Physics and Nuclear Physics \& Radiation Technology Institute (NuRI), 
Pusan National University, Busan 609-735, Republic of Korea} 
\author{Hyun-Chul Kim} 
\email{hchkim@pusan.ac.kr} 
\affiliation{Department of 
Physics and Nuclear Physics \& Radiation Technology Institute (NuRI), 
Pusan National University, Busan 609-735, Republic of Korea} 
\date{September, 2006} 
\begin{abstract}  
We investigate the leading-twist light-cone distribution amplitudes 
for the pion and kaon based on the {\em gauge-invariant} nonlocal 
chiral quark model from the instanton vacuum in the presence of 
external axial-vector currents.  We find that the nonlocal 
contribution from the gauge invariance has much effects on the pion 
distribution amplitudes, while it changes mildly the kaon ones. 
We also study the Gegenbauer moments of the distribution amplitudes 
and compare them with the empirical analysis of the CLEO data. 
 
\end{abstract} 
 
\maketitle 
\section{Introduction} 
The meson light-cone distribution amplitude (DA) provides essential 
information on the nonperturbative structure of mesons.  In 
particular, the leading-twist meson DAs play a role of input for 
describing hard exclusive reactions due to factorization 
theorems~\cite{Efremov:1979qk,Lepage:1979zb,Lepage:1980fj, 
Chernyak:1983ej}.  In particular, the pion DA has been investigated            
extensively in various theoretical approaches: For example, in the QCD 
sum rules (QCDSR)~\cite{Chernyak:1983ej,Braun:1988qv,Bakulev:1994su, 
Bakulev:2001pa,Bakulev:2005vw,Ball:2005ei}, in lattice QCD 
(LQCD)~\cite{DelDebbio:2002mq,Dalley:2002nj}, in the chiral quark 
model ($\chi$QM) from the instanton 
vacuum~\cite{Petrov:1998kg,Dorokhov:2002iu,Nam:2006au}, 
in the NJL models~\cite{Praszalowicz:2001wy,Praszalowicz:2001pi, 
RuizArriola:2002bp} and so on.  
Recently, the CLEO experiment~\cite{Gronberg:1997fj} has measured the 
$\gamma^*\gamma\to\pi$ transition form factor, which gives criteria 
for judging the existing theoretical pion
DAs~\cite{Schmedding:1999ap}.  Bakulev \textit{et 
al.}~\cite{Bakulev:2002uc,Bakulev:2003cs} have carried out similar
analyses.   
 
In the previous work~\cite{Nam:2006au}, we have investigated the 
leading-twist pion and kaon DAs within the framework of the nonlocal 
$\chi$QM from the instanton vacuum with SU(3) symmetry breaking 
explicitly taken into account.  However, it is well known that the 
nonlocal interaction breaks the conservation of the vector and 
axial-vector currents~\cite{Chretien:1954we,Holdom,Pobylitsa:1989uq}. 
The nonlocal $\chi$QM from the instanton vacuum suffers from the same 
difficulty due to the zero-mode approximation~\cite{Pobylitsa:1989uq}.    
Since the leading-twist meson DAs involve the axial-vector operator, 
one has to deal with the problem of the current conservation.  
While Ref.~\cite{Pobylitsa:1989uq} proposed a systematic way to remedy 
this problem of the current conservation, one has to handle the 
integral equation.  Refs.~\cite{Musakhanov:2002xa,Kim:2004hd} derived 
the light-quark partition function in the presence of the external 
gauge fields.  With this gauged partition function, it was shown that 
the low-energy theorem for the transition from two-photon state to the 
vacuum via the axial anomaly was satisfied~\cite{Musakhanov:2002xa}. 
Moreover, the magnetic susceptibility of the QCD vacuum was properly 
obtained~\cite{Kim:2004hd}.  
 
The conservation of the axial-vector current concerning the 
leading-twist pion DA was already discussed in 
Refs.~\cite{Anikin:2000bn,RuizArriola:2002bp} by using the 
instanton-motivated separable nonlocal kernel for the effective action 
and the NJL model with the Pauli-Villars regularization, respectively. 
It was pointed out that the nonlocal contributions are about $30\%$ of 
the local one~\cite{Anikin:2000bn}.  In the present work, we want to 
investigate the leading-twist pion and kaon DAs, employing 
the method developed in Refs.~\cite{Musakhanov:2002xa,Kim:2004hd} 
with the conservation of the axial-vector current considered.   
We will show that the nonlocal part of the pion DAs arising from the 
current conservation plays an important role in improving the 
results of the corresponding Gegenbauer moments. 
 
The present work is organized as follows:  In Section II we show how 
to make the nonlocal effective chiral action gauge-invariant from the 
instanton  vacuum.  In Section III we show how to calculate the 
leading-twist pion and kaon distribution amplitudes in the present 
approach.  In Section IV we discuss numerical results.  In Section V  
we summarize the present work and draw conclusions. 
\section{Gauge-invariant effective chiral action} 
In this Section, we briefly review how one can make the low-energy 
effective QCD partition function from the instanton vacuum in the 
presence of external axial-vector fields, following the approach 
developed in Ref.~\cite{Musakhanov:2002xa,Kim:2004hd}.  The partition 
function derived from the instanton 
vacuum~\cite{Shuryak:1981ff,Diakonov:1985eg} can be written in 
Euclidean space:   
\begin{eqnarray}  
\mathcal{Z}&=& \int {\cal D} \psi {\cal D} \psi^\dagger 
{\cal D} {\mathcal{M}} 
\exp\int d^4 x\Big[ \psi^{\dagger}_{f} (x) 
(i\rlap{/}{\partial}+im_f)\psi_{f}(x)\nonumber 
\\&&\hspace{2cm}+i\int\frac{d^4k\,d^4p}{(2\pi)^8} e^{-i(k-p)\cdot x} 
\psi^{\dagger}_{f}(k)\sqrt{M_f(k)}U^{\gamma_5}_{fg}\sqrt{M_g(p)}\psi_{g}(p) 
\Big],   
\label{eq:PF} 
\end{eqnarray}  
where $\psi_f$, $m_f$, and $M_f$ denote the quark fields, the current 
quark mass, and the momentum-dependent dynamical quark mass with 
flavor $f$, respectively.  The $U^{\gamma_5}$ or $\mathcal{M}$ 
represents the pseudo-Goldstone field defined as follows:    
\begin{eqnarray} 
U^{\gamma_5}=U(x) \frac{1+\gamma_5}{2} + U^\dagger (x) 
\frac{1-\gamma_5}{2}=1+\frac{i}{F_\mathcal{M}}\gamma_5 
\mathcal{M}^a\lambda^a-\frac{1}{2F^2_\mathcal{M}}\mathcal{M}^2\cdots,    
\label{ufield} 
\end{eqnarray}  
where $\lambda$ stands for the $3\times3$ Gell-Mann matrix.   
Musakhanov~\cite{Musakhanov:1998wp,Musakhanov:vu} 
improved the partition function in Eq.~(\ref{eq:PF}) by taking into 
account effects of flavor SU(3) symmetry breaking, so that the 
dynamical quark mass acquires the contribution of the $m_f$ 
corrections: 
\begin{equation} 
M_{f}(k)=M_{0}F^2(k)\left[ 
  \sqrt{1+\frac{m_{f}^{2}}{d^{2}}}-\frac{m_{f}}{d}\right], 
\label{dqm}    
\end{equation} 
where $M_0$ is the dynamical quark mass with zero momentum transfer in 
the chiral limit.  Its value is determined by the saddle-point 
equation: $M_0\simeq350$ MeV. $F(k)$ is the momentum-dependent part 
which arises from the Fourier transform of the fermionic zero-mode 
solutions in the instantons.  Note that we parameterize $M_0(k)$ as a 
simple-pole type to make the analytic continuation to Minkowski 
space easy.    
 
The momentum-dependent dynamical quark mass $M_f(p)$ breaks the 
conservation of the N\"other currents, i.e. it violates Ward-Takahashi  
identities.  We want to show how to remedy this problem in the 
following.  We first define the total quark propagator $\tilde S $ in the 
presence of the instanton ensemble $A$ and external axial-vector 
field  $a_\mu=a_\mu^a\lambda^a/2$ and the quark propagator $\tilde 
S_i$ with a single instanton $A_i$ as well as the $a_\mu$: 
\begin{equation} 
\tilde S = \frac{1}{i\rlap{/}{\partial} + g \rlap{/}{A} 
  +\rlap{/}{a}\gamma_5 +i\hat{m}},\;\;\; 
\tilde S_i=  \frac{1}{i\rlap{/}{\partial} + g \rlap{/}{A_i}  
+\rlap{/}{a}\gamma_5 +i\hat{m}}, 
\end{equation} 
where $\hat{m}$ is the current quark mass matrix: ${\rm diag}(m_{\rm 
  u}, m_{\rm d}, m_{\rm s})$.  We assume that the 
total instanton field $A$ may be approximated as a sum of the single 
instanton fields, $A=\sum_{i=1}^{N}A_i$, which is justified with the 
average size of instantons $\rho\approx 1/3 \,{\rm fm}$ and average 
inter-instanton distance $R \approx 1 \, {\rm 
  fm}$~\cite{Shuryak:1981ff,Diakonov:1985eg}.  
Defining the quark propagator $\tilde S_0$ with external fields 
$a_\mu$ and the free one $S_0$ as follows: 
\begin{equation} 
\tilde S_0=\frac{1}{i\rlap{/}{\partial} + \rlap{/}{a}\gamma_5 
  +i\hat{m}} ,\,\,\, S_0=\frac{1}{i\rlap{/}{\partial} +i\hat{m}}, 
\end{equation} 
the $\tilde S$ can be expanded with respect to a single instanton: 
\begin{equation} 
\label{S-tot} 
\tilde S=\tilde S_0+\sum_i (\tilde S_i-\tilde 
S_0)+\sum_{i\not=j} (\tilde S_i-\tilde S_0)\tilde S^{-1}_0(\tilde 
S_j-\tilde S_0)+\cdots . 
\end{equation} 
In order to specify the gauge dependence, we rewrite $\tilde S_i$ and 
$S_0$ in the following form: 
\begin{eqnarray} 
\tilde S_i&=&L_iS'_{i}L^{-1}_i,\;\; S'_{i}=\frac{1}{i\rlap{/}{\partial} 
+g\rlap{/}{A_i} +\rlap{/}{a'}\gamma_5+i\hat{m}},\cr 
\tilde S_0&=&L_iS'_{0}L^{-1}_i,\;\; 
S'_{0i}=\frac{1}{i\rlap{/}{\partial} + \rlap{/}{a'}\gamma_5 + 
i\hat{m}}, 
\end{eqnarray} 
where 
\begin{equation} 
\rlap{/}{a^\prime}\gamma_5=L^{-1}(i\rlap{/}{\partial} 
+\rlap{/}{a}\gamma_5)L. 
\end{equation} 
The $L$ denotes the Wilson gauge connection which is expressed as the 
path-ordered exponent: 
\begin{equation} 
L_i(x,z_i)={\rm P} \exp\left(i\int_{z_i}^x d\xi_\mu 
  a_\mu(\xi)\gamma_5\right), 
\label{transporter} 
\end{equation} 
where $z_i$ denotes an instanton position.  Having carried out the 
manipulation as in Refs.~\cite{Musakhanov:2002xa,Kim:2004hd}, we 
arrive at the low-frequency part of the fermionic determinant:   
\begin{eqnarray} 
  \label{part-func} 
 &&\tilde{\rm Det}_{\mathrm{low}} = 
 \left(\det(i\rlap{/}{\partial} +\rlap{/}{a}\gamma_5 + 
 i\hat{m})\right)^{-1} \int \prod_{f}D\psi_f D\psi^{\dagger}_{f} 
\\\nonumber 
&\times& 
 {\rm e}^{\left(\int d^4 x 
\psi_{f}^{\dagger} (i\rlap{/}{\partial} \,+\,\rlap{/}{a}\gamma_5 
 \, +\, im_f )\psi_{f}\right)} 
 \prod_{f}\left\{\prod_{+}^{N_{+}} 
V_{+,f}[\psi_{f}^{\dagger},\psi_f ] 
\prod_{-}^{N_{-}}V_{-,f}[\psi_{f}^{\dagger},\psi_f ]\right\}\; ,  
\end{eqnarray} 
where 
\begin{eqnarray} 
  \label{tildeV} 
\tilde V_{\pm,f}[\psi_{f}^{\dagger},\psi_f ]=\int d^4 x 
\left(\psi_{f}^{\dagger} (x)\,L(x,z) i\rlap{/}{\partial} 
\Phi_{\pm, 0} (x; \xi_{\pm})\right) \int d^4 y \left(\Phi_{\pm , 
0}^\dagger (y; \xi_{\pm} ) (i\rlap{/}{\partial} L (y,z) 
\psi_{f} (y)\right) 
\end{eqnarray} 
with the fermionic zero-mode solutions $\Phi_{\pm,0}$. 
 
The gauge connection $L$ has some arbitrariness 
due to its path dependence.  However, we can show that such 
dependence can be minimized.  To be more specific, we consider the 
extended zero mode: 
\begin{equation} 
 \label{eq:extended} 
(i\rlap{/}{\partial} + \rlap{/}{A}+\rlap{/}{a}\gamma_5)| 
\tilde\Phi_0\rangle = 0,\,\,\, 
|\tilde\Phi^{(1)}_0\rangle \simeq |\Phi_0\rangle - 
S_{NZ}\rlap{/}{a}\gamma_5 |  \Phi_0\rangle,  
\end{equation} 
where $A$ denotes an instanton field located at $z$, 
$|\tilde\Phi^{(1)}_0\rangle$ stands for the solution in the presence 
of the external field, and $S_{NZ}$ the well-known non-zero mode of 
the propagator in the instanton field (see the review 
\cite{Schafer:1996wv} and references therein).  Here, we have assumed 
tacitly that the momentum of the external field is small.  Inserting 
the gauge connection $L(x,z)$ into Eq.~(\ref{eq:extended}), we obtain: 
\begin{equation} 
(i\rlap{/}{\partial} + \rlap{/}{A}+\rlap{/}{a}\gamma_5)|\Phi_0'\rangle 
= 0, \,\,\, a_{\mu}'\gamma_5=L^{-1}(i\partial_{\mu}+a_{\mu}\gamma_5)L,   
\,\,\,|\Phi_0'\rangle = L^{-1} (x,z) |\tilde\Phi_0\rangle. 
\label{eq:path1} 
\end{equation} 
If we utilize the relations $(i\rlap{/}{\partial} + 
\rlap{/}{A})|\Phi_0\rangle=0$ and $S_{NZ}\,(i\rlap{/}{\partial} + 
\rlap{/}{A})=1-|\Phi_0\rangle\langle\Phi_0|$, the solution 
$|\Phi^{(1)'}_0\rangle$ in Eq.~(\ref{eq:path1}) can be reduced to the 
corresponding solution $|\tilde\Phi^{(1)}_0\rangle$ without any problem 
of the path dependence arising from the gauge connection $L$. 
 
However, if the zero-mode approximation $S_{NZ}\approx 
\frac{1}{i\rlap{/}{\partial}}=S_{00}$ is used to find the solution 
\footnote{In fact, the zero-mode approximation is the origin of the 
problem of the current non-conservation from the instanton vacuum. 
Pobylitsa suggested a more reliable approximation to restore the 
current conservation~\cite{Pobylitsa:1989uq}.} 
\begin{equation} 
|\Phi^{(1)'}_{00}\rangle = |\Phi_0\rangle - 
S_{00}\rlap{/}{a}'\gamma_5| \Phi_0\rangle, 
\label{eq:phi00} 
\end{equation} 
then  
\begin{equation} 
|\tilde\Phi^{(1)}_{00}\rangle= |\Phi_0\rangle -S_{00} 
\rlap{/}{a}\gamma_5 | \Phi_0\rangle + S_{00} \left(i\int 
d\xi_{\mu}a_{\mu}(\xi)\gamma_5\right) | \Phi_0\rangle 
\label{12} 
\end{equation} 
with applying the inverse gauge connection $L^{-1}$ to 
Eq.~(\ref{eq:phi00}).  We can easily see that Eq.~(\ref{12})   
depends on the path of the gauge connection.  We need to choose the 
path in such a way that $\| 
\tilde\Phi^{(1)}_0-\tilde\Phi^{(1)}_{00}\|^2$ can be minimized.  As 
shown explicitly in Ref.~\cite{Kim:2004hd}, the straight line 
provides the most optimized path.  
 
Having treated the gauge problem in the presence of the external 
axial-vector field, we finally obtain the low-energy partition 
function of the nonlocal $\chi$QM from the instanton vacuum: 
\begin{eqnarray}  
{\cal Z}[a,m]&=&\int{\cal D}\psi{\cal D}\psi^\dagger 
{\cal D} {\cal M} 
\exp\int d^4 x\Big[ \psi^{\dagger}_{f} (x) 
(i\rlap{/}{\partial}+\rlap{/}{a}\gamma_5+im_f)\psi_{f}(x)\nonumber 
\\&+&i\int\frac{d^4k\,d^4p}{(2\pi)^8} e^{-i(k-p)\cdot x} 
\psi^{\dagger}_{f}(k)\sqrt{M_f(k_{\mu}+a_{\mu}\gamma_5)} 
U^{\gamma_5}_{fg} 
\sqrt{M_g(p_{\mu}+a_{\mu}\gamma_5)}\psi_{g}(p)\Big]  
\label{GPF} 
\end{eqnarray} 
with the effective chiral action in the presence of the external 
axial-vector source 
\begin{eqnarray} 
\label{eq:EA} 
{\cal S}_{\rm eff}[a,m]=-N_c{\rm Tr}\ln\left[i\rlap{/}{\partial} 
+\rlap{/}{a}\gamma_5+im_f+i\sqrt{M_f(i\partial_{\mu} 
+a_{\mu}\gamma_5)} 
U^{\gamma_5}_{fg}\sqrt{M_g(i\partial_{\mu}+a_{\mu}\gamma_5)}\right], 
\end{eqnarray} 
where ${\rm Tr}$ denotes the functional trace and traces over flavor 
and spin spaces, generically. 
\section{Meson distribution amplitudes} 
The leading-twist pseudoscalar meson light-cone distribution amplitudes 
(DAs) can be defined with the conserved axial-vector current as 
follows (here, we consider the light-cone gauge 
$A\cdot n=0$):    
\begin{eqnarray} 
\langle0|\bar{q}_{f}(\tau\hat{n})\gamma_\mu\gamma_5 
q_{g}(-\tau\hat{n})|\mathcal{M}(P)\rangle 
 = i\sqrt{2}F_{\mathcal{M}}P_{\mu}\int^1_0du\,e^{i(2u-1) 
P\cdot\tau\hat{n}}\phi_{\mathcal{M}}(u), 
\label{DA} 
\end{eqnarray} 
where $\mathcal{M}$ denotes the flavor SU(3) octet pseudoscalar meson 
field with the on-mass shell momentum $P^2=m^2_{\cal{M}}$ in the   
light-cone frame. 
In the present work, we choose $\mathcal{M}=\pi^+$ or  
$\mathcal{M}=K^+$. $u$ and $\hat{n}$ stand for the longitudinal  
momentum fraction and light-like vector, respectively. $F_{\mathcal{M}}$ is 
the pseudoscalar meson decay constant. Note that, when the 
$\tau$ approaches to zero, Eq.~(\ref{DA}) becomes the pseudoscalar meson decay 
amplitude due to the DA normalization 
condition:   
\begin{eqnarray} 
\int^1_0du\,\phi_{\mathcal{M}}(u)=1.   
\label{norm}     
\end{eqnarray} 
This equation will be used for determining the scale parameter of the
model.  At this point, we want emphasize that in order to calculate
the light-cone DA, it is necessary for us to work in Minkowski space
in spite of using the instanton framework which is well-defined in
Euclidean space as discussed before. For this purpose, we assume
simple analytic continuation (the Wick-rotation) between the two
spaces in a practical point of view~\cite{Petrov:1998kg,Anikin:2000bn,
Praszalowicz:2001wy,Dorokhov:2002iu}.  
 
The expression for the leading-twist pion and kaon DAs in the present 
approach is obtained as follows: 
\begin{eqnarray} 
&&\phi_{\mathcal{M}}(u)=i\frac{N_c\hat{n}^{\mu}}{2F^2_\mathcal{M}} 
\int\frac{d^4k}{(2\pi)^4}\delta[uP\cdot\hat{n}-k\cdot\hat{n}]\;\; 
{\rm tr}_{\gamma}\left[\frac{\sqrt{M_f(k)}}{D_f(k)}\gamma_{\mu} 
\gamma_5\frac{\sqrt{M_g(k-P)}}{D_f(k-P)}\gamma_5\right.\cr 
&&\left.+\;\frac{\sqrt{M_f(k)_{\mu}} 
\sqrt{M_g(k-P)}}{D_f(k)}\;-\;\frac{\sqrt{M_f(k)} 
\sqrt{M_g(k-P)_{\mu}}}{D_f(k-P)}\right],\nonumber\\ 
\label{DA2} 
\end{eqnarray} 
where $\sqrt{M_f(k)_{\mu}}=\partial\sqrt{M_f(k)}/\partial k_{\mu}$. 
$\mathrm{tr}_{\gamma}$ denotes the trace over the Dirac spin 
space.  Note that the last two terms that we call the nonlocal 
contributions in the bracket make the present result gauge-invariant. 
The quark propagator $D_f(k)$ is expressed in terms of the 
momentum-dependent dynamical quark mass $M_f(k)$ and current quark 
mass $m_f$ as follows~\cite{Musakhanov:vu}:  
\begin{eqnarray} 
D_f(k)=\rlap{/}{k}-[m_f+M_f(k)]. 
\label{eq:qp} 
\end{eqnarray} 
We employ a simple-pole type parameterization for $F(k)$ in 
Eq.~(\ref{dqm}): 
\begin{eqnarray} 
F(k) = \left[\frac{n\Lambda^2}{(n\Lambda^2-k^2+i\epsilon)}\right]^{n}, 
\label{dynamicalmass} 
\end{eqnarray} 
since it is easy for the analytic continuation to Minkowski space. 
  
The derivative of the dynamical quark mass $\sqrt{M}_{\mu}$ appearing in the 
nonlocal contributions can be evaluated analytically with the simple-pole type 
form factor of Eq.~(\ref{dynamicalmass}):  
\begin{eqnarray} 
\sqrt{M_f(k)_{\mu}}=\frac{1}{2\sqrt{M_f(k)}}\frac{\partial{M_f(k)}}{\partial 
  k_{\mu}}=\frac{2n\sqrt{M_f(k)}}{(n\Lambda^2-k^2)}k_{\mu}. 
\label{derivative} 
\end{eqnarray} 
Inserting Eqs.~(\ref{dynamicalmass}) and (\ref{derivative}) into 
Eq.~(\ref{DA2}), we finally obtain the expression for the 
leading-twist light-cone pion and kaon DAs:   
\begin{eqnarray} 
\phi_{\mathcal{M}}(u)&=&i\frac{N_c}{2F^2_\mathcal{M}}\int 
\frac{d^4k}{(2\pi)^4}\delta[uP_+-k_+]\sqrt{M_f(k)}\sqrt{M_f(k-P)}\;\; 
{\rm tr}_{\gamma} \left[\frac{1}{D_f(k)}\rlap{/}{\hat{n}} 
\gamma_5\frac{1}{D_g(k-P)}\gamma_5\right.\cr 
&+&\left. \frac{2n k_+}{D_f(k)(n\Lambda^2-k^2)}-\frac{2n(k_+-P_+)}{ 
D_g(k-P)[n\Lambda^2-(k-P)^2]}\right]\cr 
&=&\frac{N_c}{2F^2_\mathcal{M}}\int\frac{dk_-dk^2_T}{(2\pi)^3} 
\left[\mathcal{T}_L+\mathcal{T}_{NL}\right], 
\label{DA3} 
\end{eqnarray} 
where $p_+$ stands for $p\cdot \hat{n}$.  The subscripts $L$ and $NL$ 
stand for the local and nonlocal contributions, respectively. Further 
evaluation of $\mathcal{T}_{L,NL}$ can be found in Appendix.    
\section{Results and Discussion} 
In this Section, we discuss the numerical results for the leading-twist 
light-cone pion and kaon DAs with the gauged nonlocal effective chiral 
action.  We first fix the scale parameter of the model, i.e. $\Lambda$ 
in Eq.~(\ref{dynamicalmass}).  In Ref.~\cite{Nam:2006au}, the 
$\Lambda$ was chosen to be 1.2 GeV by using the normalization 
condition in Eq.~(\ref{norm}), with which the pion and kaon decay 
constants are simultaneously reproduced.  However, since the 
Ward-Takahashi identity for the axial-vector current is broken, the 
improper Pagels-Stokar (PS) expression for the pion decay constant had 
to be used to normalize the DAs properly.  In the present 
gauge-invariant approach, we are able to determine the scale parameter 
$\Lambda$ without any trouble.  
          
In Table~\ref{table1}, we list the pion and kaon decay constants 
evaluated by using the normalization condition in Eq.~(\ref{norm}). 
As shown in Table~\ref{table1}, when we set $\Lambda\simeq1.0$ GeV, 
the values of $F_{\pi}$ are in accord with the empirical one 
($F_{\pi}=93$ MeV) for powers of the form factor $n=1,2,3$.  We find 
that without the nonlocal contribution from gauge invariance,  
the results are underestimated by about $10\sim20\%$.  The kaon decay 
constant $F_K$ is found to be about $10\,\%$ smaller than the 
empirical data $F_K=113$ MeV.  However, it is well known that the 
meson-loop corrections play a crucial role in describing the 
$F_K$~\cite{Gasser:1984gg}.  
\begin{table}[ht] 
\begin{tabular}{c|ccc||ccc} 
\hline 
&\multicolumn{3}{c||}{$F_{\pi}$ (Exp: $93$ MeV)}&\multicolumn{3}{c}{$F_{K}$ 
  (Exp: $113$ MeV)}\\  
\hline 
$n$&Local&Nonlocal&Total&Local&Nonlocal&Total\\ 
\hline 
1&87.03&28.75&91.66&92.59&31.94&97.95\\ 
2&84.47&30.38&89.77&90.87&34.37&97.15\\ 
3&83.68&31.02&89.24&90.31&35.31&96.97\\ 
\hline 
\end{tabular} 
\caption{The results of the pion and kaon decay constants $F_{\pi}$ 
and $F_K$ [MeV].}  
\label{table1} 
\end{table} 
 
\begin{figure}[t] 
\begin{tabular}{ccc} 
\includegraphics[width=5.5cm]{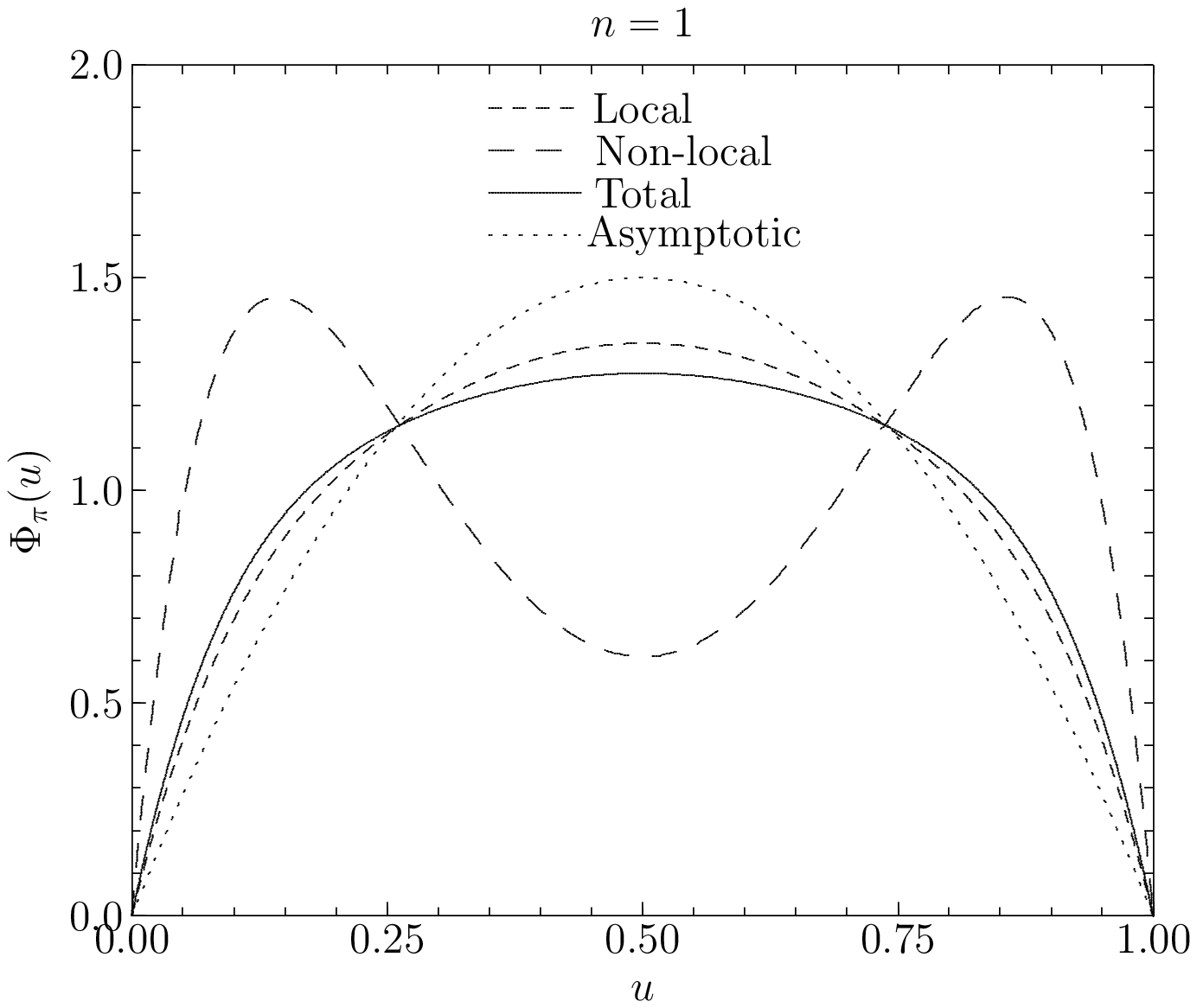} 
\includegraphics[width=5.5cm]{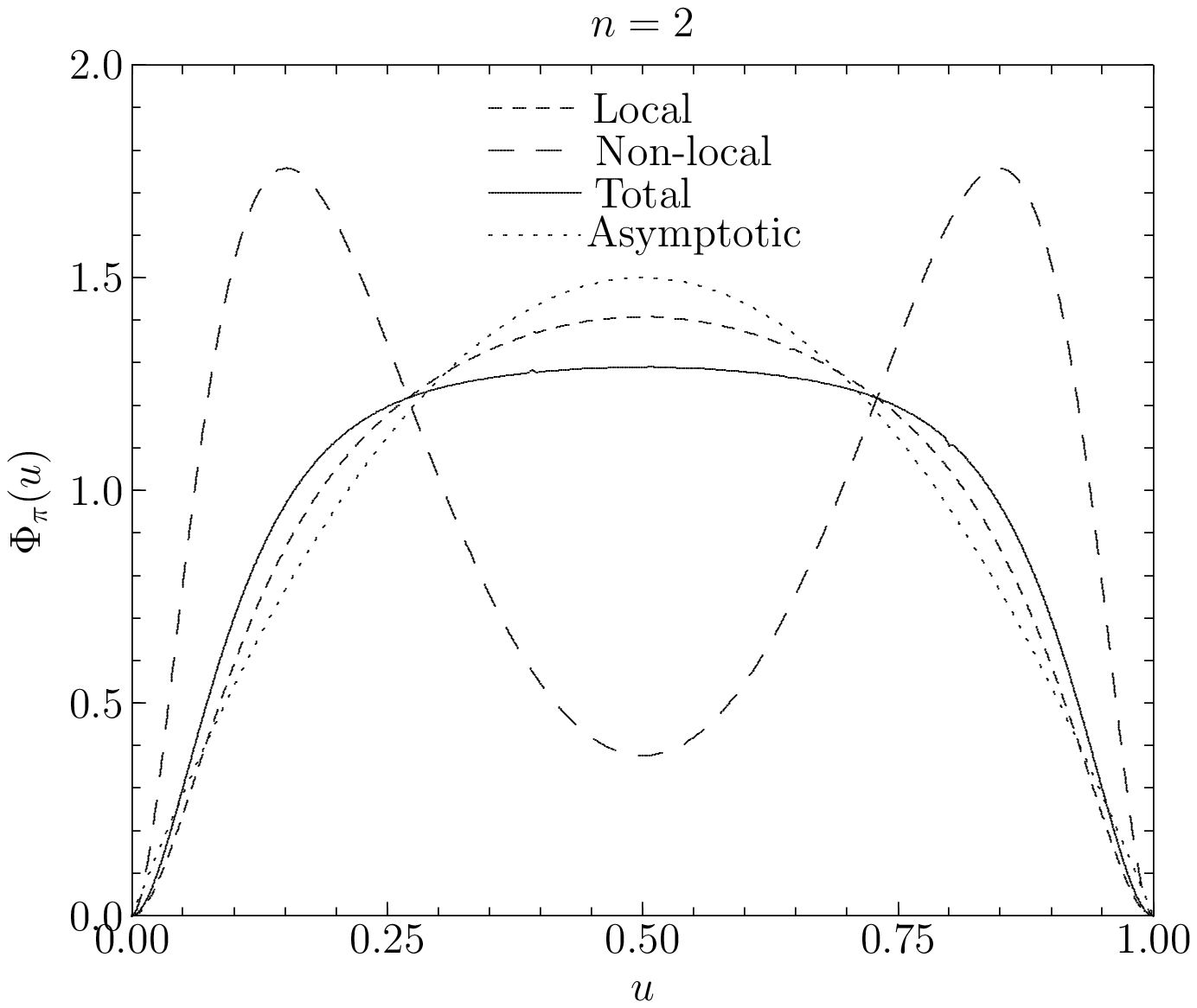} 
\includegraphics[width=5.5cm]{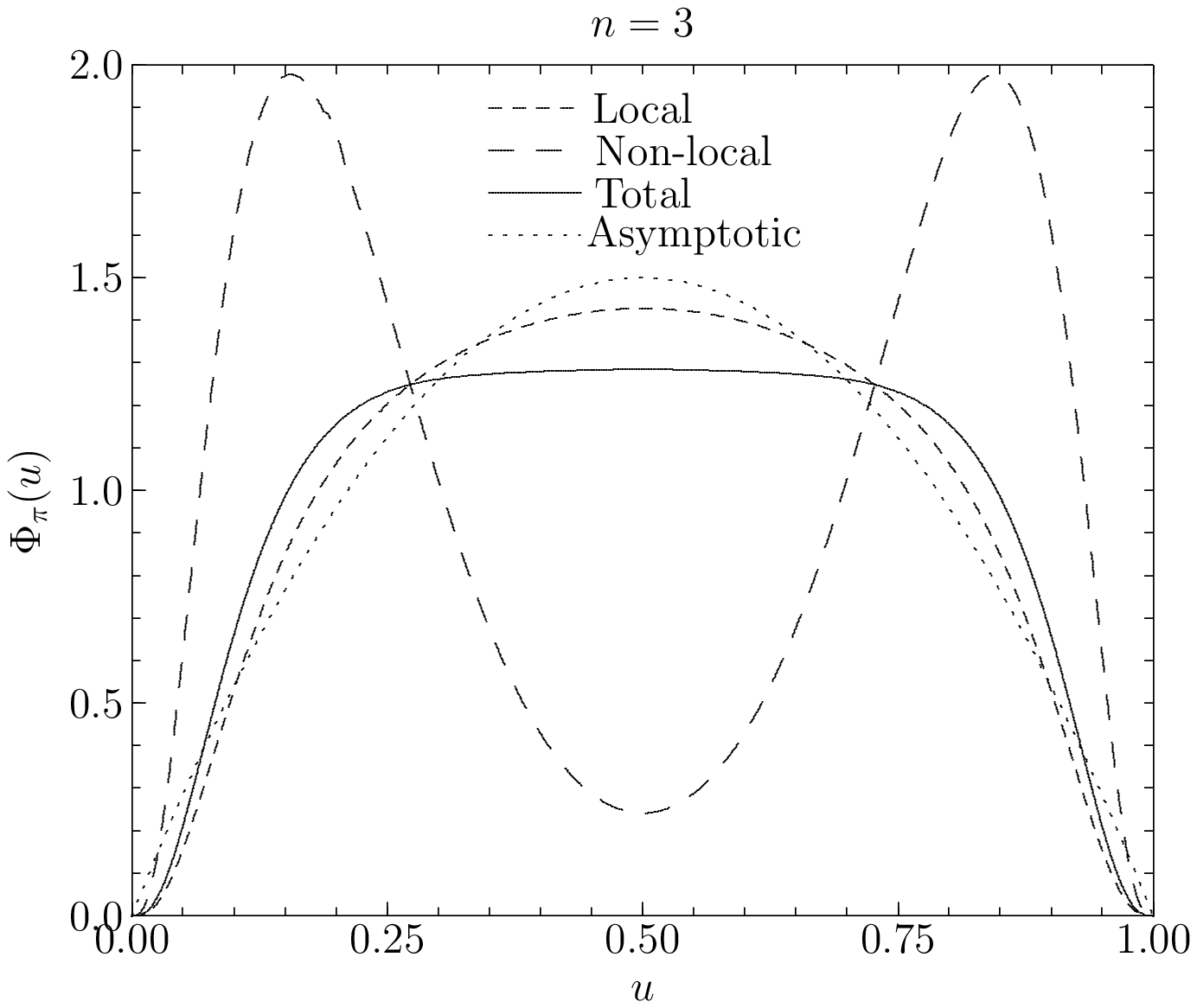} 
\end{tabular} 
\begin{tabular}{ccc} 
\includegraphics[width=5.5cm]{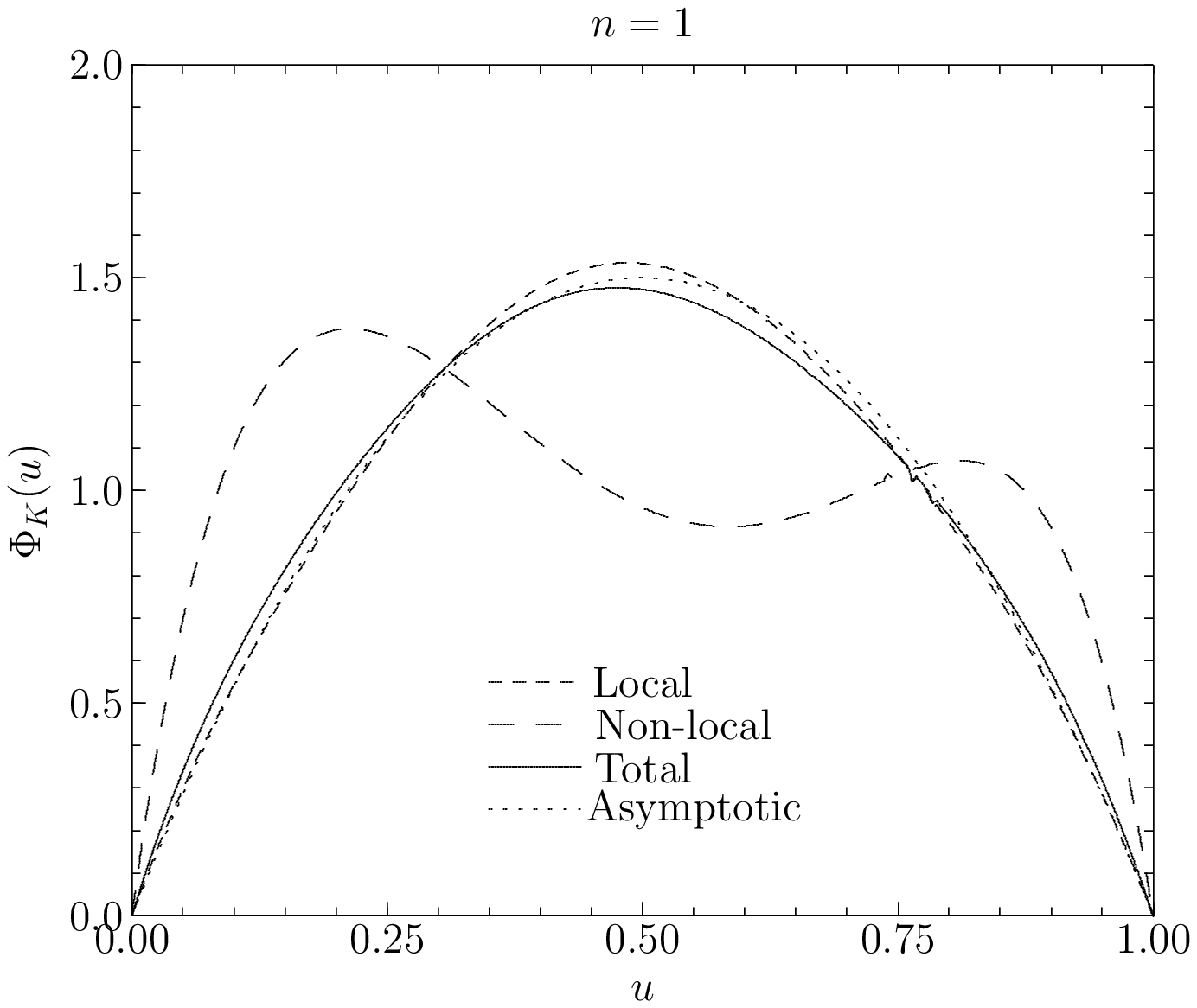} 
\includegraphics[width=5.5cm]{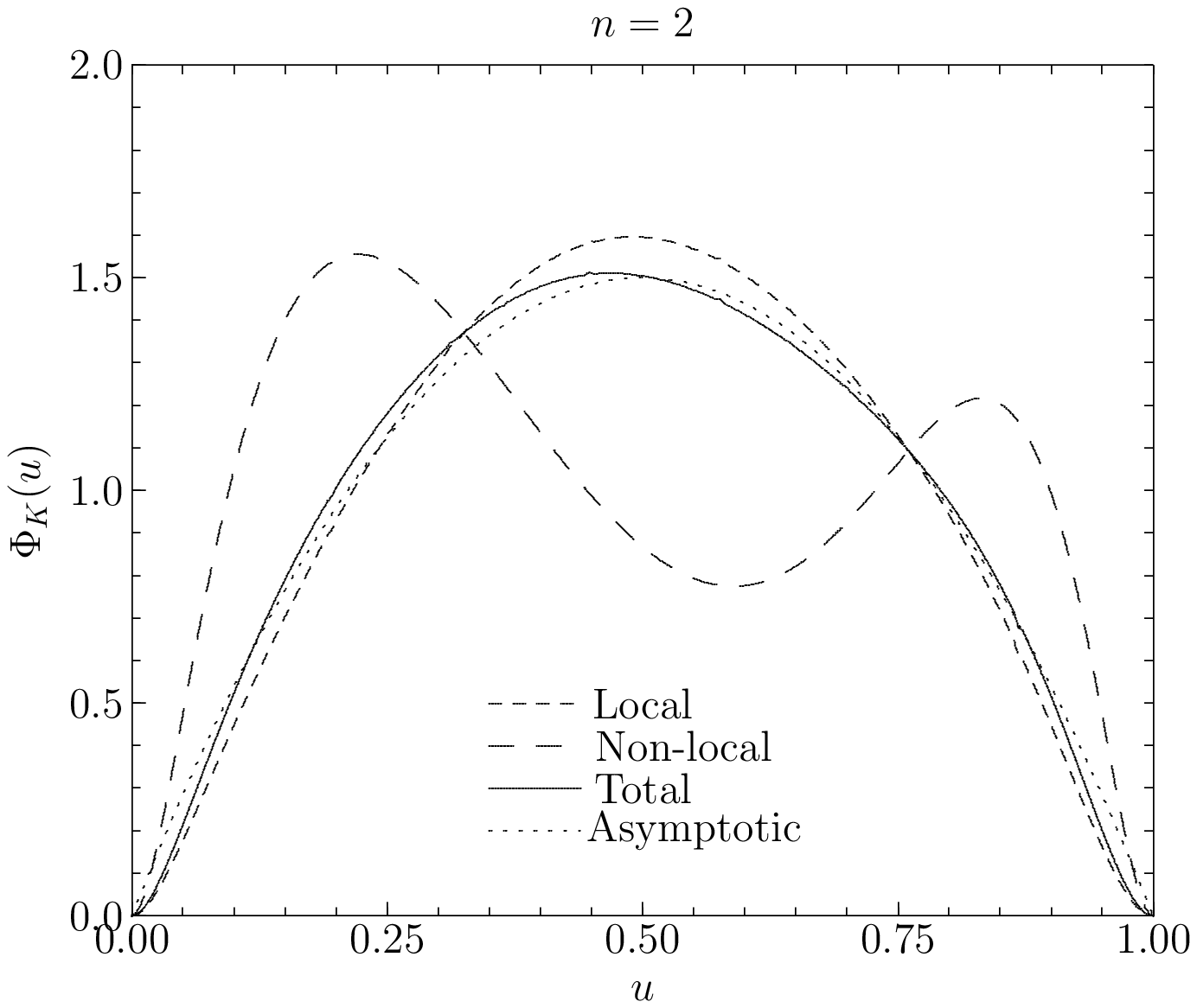} 
\includegraphics[width=5.5cm]{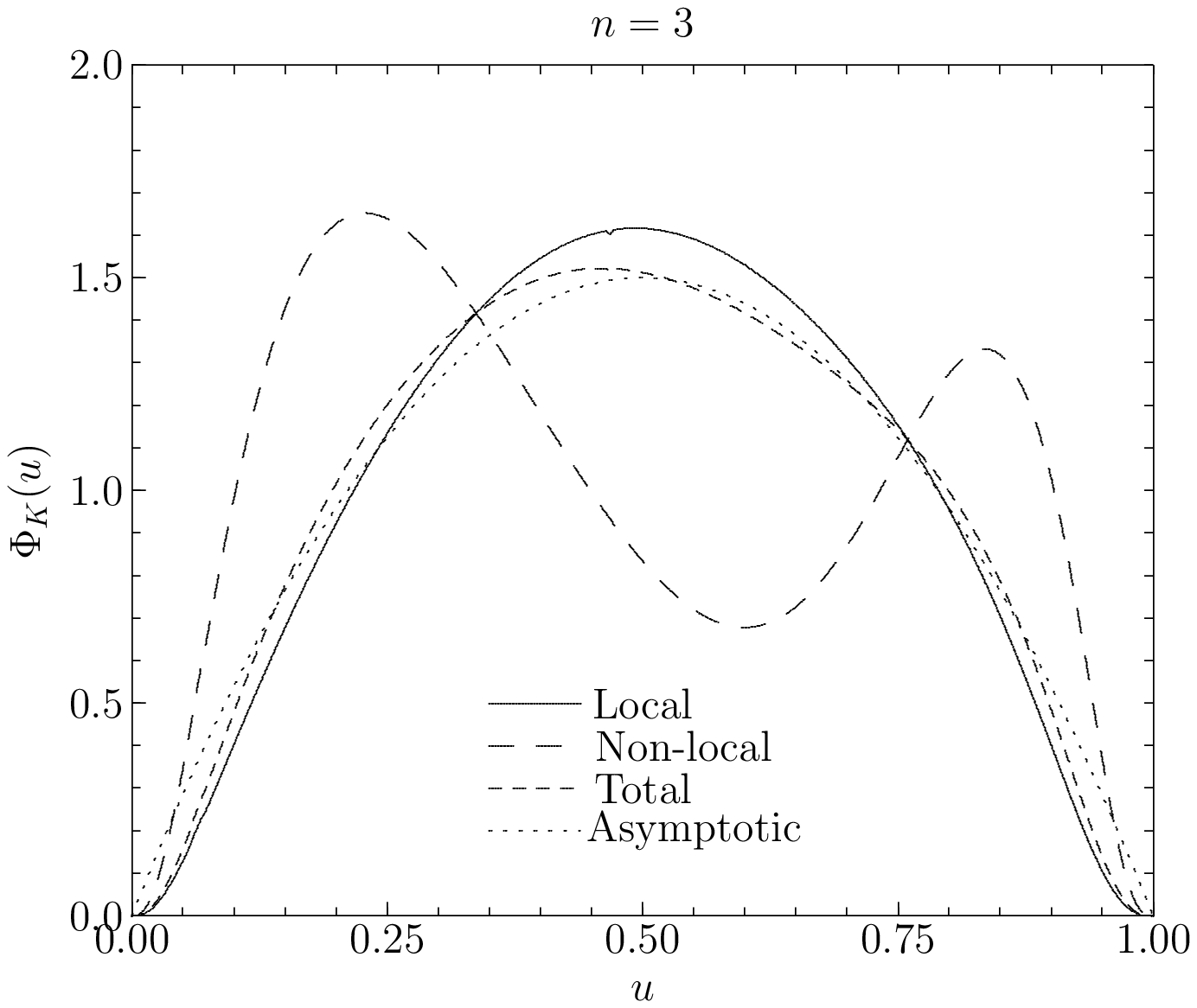} 
\end{tabular} 
\caption{Pion (upper panels) and kaon (lower panels) distribution 
amplitudes.  The dashed, long dashed, and solid curves represent the 
local, nonlocal, and total contributions, respectively.  The 
asymptotic DA is also drawn for comparison.}   
\label{fig2} 
\end{figure}

In Fig.~\ref{fig2}, we depict the pion and kaon DAs in the upper and 
lower panels, respectively.  The dashed, long dashed, and solid curves 
represent the local, nonlocal, and total contributions to the DAs, 
respectively.  For comparison, we also draw the asymptotic DA in the 
dotted curve in each panel.  As the power of the form factor $n$ 
increases, the local contributions are almost independent of $n$, 
whereas the nonlocal ones are drastically changed: Their humped shape
is getting manifest as $n$ increases.  As a result, the total pion DA  
is getting flattened in the region of $0.25\lesssim{u}\lesssim0.75$, 
as $n$ increases.  Although we do not present explicitly the pion DA 
with higher $n$, it is even more flattened, if we increase $n$ higher 
than three.  Note that the pion DA is suppressed at the end points of 
$u$.  In the lower panels of Fig.~\ref{fig2}, we find 
that while the local contribution to the kaon DA is almost symmetric, 
the nonlocal one turns out to be rather asymmetric.  However, when 
these two contributions are added, the kaon DA is unexpectedly similar 
to the asymptotic one.  However, the kaon DA is quite much suppressed 
at the end points, in particular, with higher $n$, compared to the 
asymptotic one.  It indicates that the effects of the current quark 
mass becomes obvious for the kaon DA at the end points, as shown in 
Ref.~\cite{Nam:2006au}. The suppression of the pion and kaon DAs at 
the end points corresponds to the negative second Gegenbauer moments.     
 
It is also of great use to examine the Gegenbauer polynomial to 
analyze the DAs in detail~\cite{Praszalowicz:2001wy,Nam:2006au}. The 
Gegenbauer moments for the DAs are defined by the coefficients of the 
Gegenbauer polynomial expansion of the DAs, $a_m$:  
\begin{eqnarray} 
\phi(u)&=&6u(1-u)[1+a_1C^{3/2}_1(\xi)+a_2C^{3/2}_2(\xi) 
+\cdots].  
\end{eqnarray} 
Note that the odd ($2m+1$) Gegenbauer moments are all zero for the pion 
DA because of isospin symmetry. The Gegenbauer moments to 
the sixth order ($m=6$) are evaluated and are listed in 
Table~\ref{table2} for the  pion and in Table~\ref{table3} for the 
kaon.   
\begin{table}[b] 
\begin{center} 
\begin{tabular}{c|ccc} 
\hline 
$n$&$a^{\pi}_2$&$a^{\pi}_4$&$a^{\pi}_6$\\  
\hline 
1&  0.11911 & 0.01438 & $-$0.00009\\ 
2&  0.07127& $-$0.03605 & $-$0.02103\\ 
3&  0.05339& $-$0.06088&  $-$0.02596\\ 
\hline 
\cite{Chernyak:1983ej} &$0.56$&--&--  \\ 
\cite{RuizArriola:2002bp} (2.4 GeV)&--& $0.044\pm0.016$ 
&$0.023\pm0.010$\\  
\cite{Schmedding:1999ap} (2.4 GeV)& $0.12\pm0.03$ &--&--\\  
\cite{Khodjamirian:2004ga} (1.0 GeV)& $0.26^{+0.21}_{-0.09}$ &-- 
&--\\   
\cite{Gockeler:2005jz} (2.24 GeV)& $0.236(82)$ &--&--\\ 
\cite{Ball:2006wn} (1.0 GeV)& $0.25\pm0.15$  &--&-- \\ 
\cite{Braun:1989iv} (1.0 GeV)& $0.44$  &$0.25$ &-- \\ 
\hline 
\end{tabular} 
\end{center} 
\caption{The Gegenbauer moments for the pion DA.  The numbers 
in the parentheses stand for the renormalization scale.}   
\label{table2} 
\end{table} 
 
Our results of $a^{\pi}_2$ for $n=1$ are rather consistent with 
those of various model calculations listed in Table~\ref{table2}. We  
find that the absolute values of $a^{\pi}_2$ are in general larger than 
those given in Ref.~\cite{Nam:2006au} in which the nonlocal 
contributions were not considered.  Moreover, we verified that the  
difference between those with and without the nonlocal ones 
becomes evident as $n$ increases.  We obtain the negative values of 
$a^{\pi}_4$ for $n=2$ and $3$, which are different from those of other 
models ~\cite{RuizArriola:2002bp,Braun:1989iv}.  As mentioned 
previously, the sign for $a_4$ plays an important role in 
the end-point behavior of the pion and kaon DAs.  The negative sign 
indicates that the DAs are suppressed at the end points, i.e. the  
DAs turn out to be in concave shapes at the end points.  
 
In Table~\ref{table3}, we list the Gegenbauer moments for the kaon DA, 
including both odd and even ones.  The most interesting feature shown 
here is that the signs of the calculated $a^K_1$ are all  
negative.  The results of the QCD sum rule calculations are not 
conclusive in determining the sign of 
$a^K_1$~\cite{Ball:2003sc,Braun:2004vf,Khodjamirian:2004ga,Ball:2006wn}. 
An interesting discussion on this issue can be found in 
Ref.~\cite{Khodjamirian:2004ga}.  In addition, the $a^K_2$ are also  
negative for $n=2,3$, being obviously different from the pion.  Note 
that the values of $a^K_2$ are all positive for the QCDSR calculations 
in Refs.~\cite{Gockeler:2005jz,Ball:2006wn,Braun:1989iv}.    
\begin{table}[b] 
\begin{center} 
\begin{tabular}{c|cccccc} 
\hline 
$n$&$a^K_1$&$a^K_2$&$a^K_3$&$a^K_4$&$a^K_5$&$a^K_6$\\  
\hline 
1&  $-$0.01879 &  0.02525 &  0.00467 &  0.00676 & $-$0.00283 &  0.00069\\  
2& $-$0.01193 & $-$0.01935 &  0.00599&  $-$0.02470&  $-$0.00371&  $-$0.01183\\  
3& $-$0.01012 & $-$0.03694 &  0.00706&  $-$0.03840&  $-$0.00566&  $-$0.01365\\  
\hline 
\cite{Khodjamirian:2004ga} (1.0 GeV)&$0.07^{+0.02}_{-0.03}$ 
&$0.27^{+0.37}_{-0.12}$ &-- &-- &-- &--  \\  
\cite{Ball:2006wn} (1.0 GeV)&&$0.30\pm0.15$ &-- &-- &-- &-- \\ 
\cite{Ball:2003sc} (1.0 GeV) &$-0.18\pm0.09$ &$0.16\pm1.10$ &-- &-- &-- 
&--\\  
\cite{Braun:2004vf} (1.0 GeV)&$0.10\pm0.12$&-- &-- &-- &-- &--\\  
\hline 
\end{tabular} 
\end{center} 
\caption{The Gegenbauer moments for the kaon DA.  The numbers 
in the parentheses stand for the renormalization scale.}  
\label{table3} 
\end{table} 
 
\begin{figure}[t] 
\begin{tabular}{c} 
\includegraphics[width=10cm]{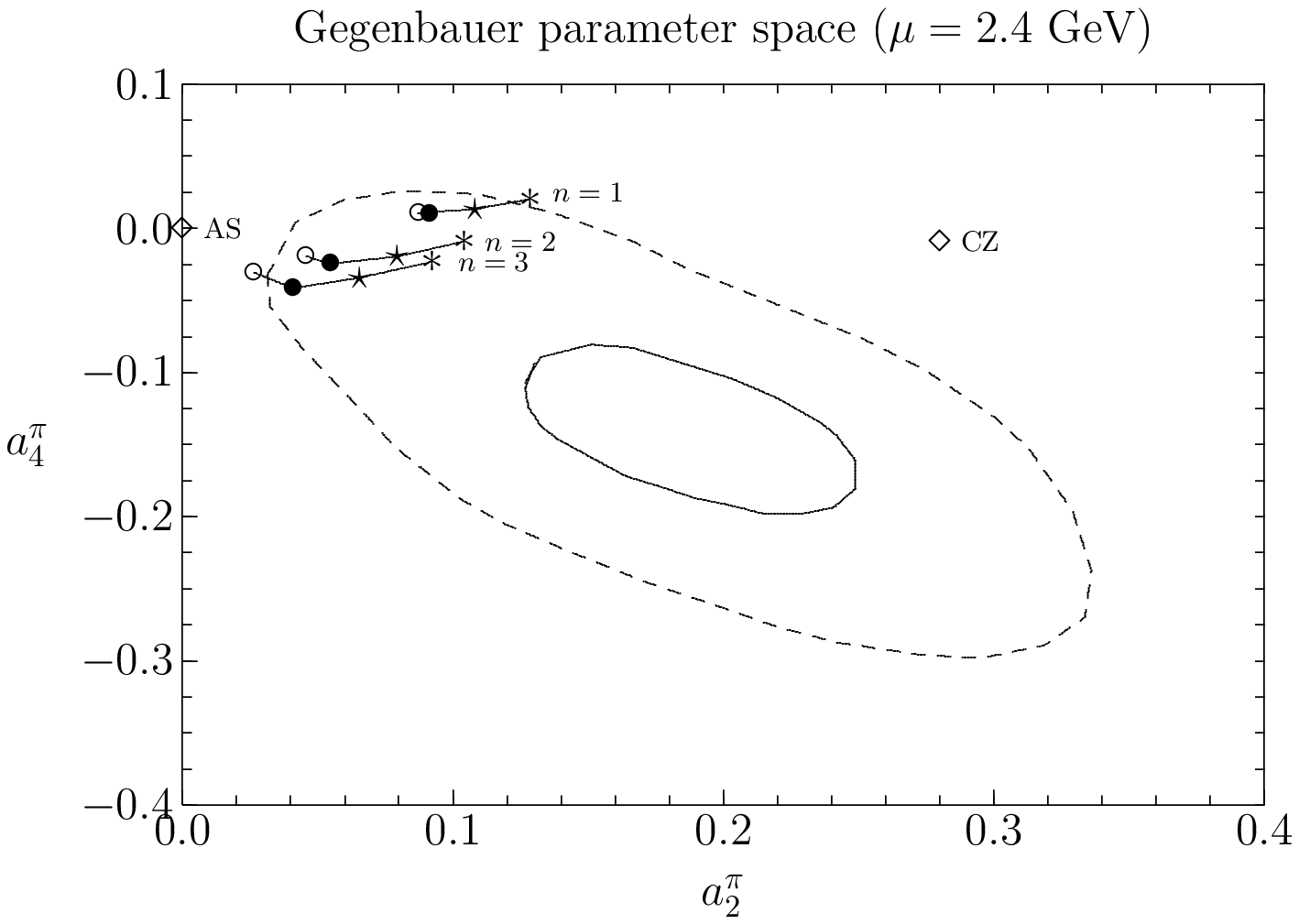} 
\end{tabular} 
\caption{Gegenbauer parameter space ($a_2-a_4$) for the pion DA   
with the CLEO experimental data analysis taken from 
Ref.~\cite{Schmedding:1999ap,Gronberg:1997fj}  
in the renormalization scale $\mu=2.4 $GeV. The solid and dashed 
ellipses indicate the $68\%\,(1\sigma)$ and $95\%\,(2\sigma)$ confidential 
levels of the analysis.  The open circles ($\circ$) denote  
the results of the Gegenbauer moments without the nonlocal 
contributions ($\Lambda=1.2$ GeV)~\cite{Nam:2006au}.  The close 
circles ($\bullet$), stars ($\star$), and asterisks ($\ast$)  
stand for the cases with the nonlocal contributions 
for $\Lambda=1.0$ GeV, $1.2$ GeV and $1.5$ GeV, respectively.  The 
results with the same power $n$ and different $\Lambda$ are connected 
by the solid line.}    
\label{fig3} 
\end{figure}

The analysis of the Gegenbauer moments $a^{\pi}_2$ and $a^{\pi}_4$ has 
been performed first by Schmedding and 
Yakovlev~\cite{Schmedding:1999ap}, based on the CLEO data of  
the transition form factor $F_{\gamma^*\gamma\pi}(Q^2)$ at the 
renormalization point $\mu=2.4$ GeV (SY-scale).  In order to compare 
them to the present results of the Gegenbauer moments, it is necessary
to evolve the scale of the present calculation to the SY-scale by
using the one-loop QCD evolution equation.  Note that the anomalous  
dimension for the Gegenbauer moments for the leading-twist DAs are 
obtained as follows~\cite{Efremov:1979qk,Lepage:1979zb,Lepage:1980fj, 
Chernyak:1983ej}:    
\begin{eqnarray} 
\gamma^{(0)}_m=-\frac{8}{3}\left[3+\frac{2}{(m+1)(m+2)}-4\sum_{m'=1}^{m+1} 
\frac{1}{m'}\right],  
\end{eqnarray}    
where $\gamma^{(0)}_1=7.11$, $\gamma^{(0)}_2=11.11$, $\gamma^{(0)}_3=13.95$, 
$\gamma^{(0)}_4=16.18$. Thus, the Gegenbauer moments in two different 
renormalization scales, $\Lambda_1=2.4$ GeV and $\Lambda_2=1.0$ GeV, 
can be compared by the following relation:  
\begin{eqnarray} 
a_m(\Lambda_1)=a_m(\Lambda_2)\left[\frac{\alpha(\Lambda_1)}{\alpha(\Lambda_2)} 
\right]^{\gamma^{(0)}_m/(2\beta_0)}\sim  
a_m(\Lambda_2)\left[\frac{\ln[\Lambda_2/\Lambda_{\rm QCD}]}{\ln[ 
\Lambda_1/\Lambda_{\rm QCD}]}\right]^{\gamma^{(0)}_m/(2\beta_0)}. 
\end{eqnarray} 
Here, we use $\beta_0=9$ and $\Lambda_{\rm QCD}\simeq0.2$ GeV. The contours on 
the Gegenbauer parameter plane ($a_2,a_4$) analyzed from the CLEO experiment 
are depicted together with our results in Fig.~\ref{fig3}.    
The solid and dashed ellipses indicate the $68\%$  
and $95\%$ confidential levels, respectively~\cite{Schmedding:1999ap}. 
We also indicate the positions for the asymptotic DA (AS) and 
for the Chenyak-Zhitnitsky (CZ) DA~\cite{Chernyak:1977as}.  The closed 
circles ($\bullet$) denote the values of $(a^{\pi}_2,a^{\pi}_4)$
calculated in the present work.  On the contrary, the open circles 
($\circ$) are taken from Ref.~\cite{Nam:2006au} in which the gauge 
invariance of the effective chiral action was not considered.  We 
observe that the closed circles for $n=1,2,3$ are all inside the 
$95\%$ confidential levels.  However, when the nonlocal  
contributions are turned off (open circles), which is identical to the 
results of Ref.~\cite{Nam:2006au}, $a^{\pi}_2$ and $a^{\pi}_4$ for
$n=3$ lie outside the $95\%$ confidential regions. We also see that
the values of $|a^{\pi}_4|$ become slightly larger with the nonlocal   
contributions.  We infer from it that the pion DA is getting more 
flattened by larger values of $|a^{\pi}_4|$.   
 
In addition, we depict the results for $\Lambda=1.2$ GeV and $1.5$ GeV 
in the stars ($\star$) and asterisks ($\ast$), respectively in 
Fig.~\ref{fig3}, to test the dependence on the scale parameter of the
model $\Lambda$.  We see that as $\Lambda$ increases the positions of 
$(a^{\pi}_2,a^{\pi}_4)$ move toward the center of the $95\%$ 
confidential level.  The $a^{\pi}_4$ is less sensitive to the 
$\Lambda$, compared to $a^{\pi}_2$.  We verified that the  
calculated pion decay constants $F_{\pi}$ for $\Lambda=1.2$ GeV and 
$1.5$ GeV are only smaller by about $10\%$ than the empirical one.      
 
\begin{figure}[ht] 
\begin{tabular}{ccc} 
\includegraphics[width=5.5cm]{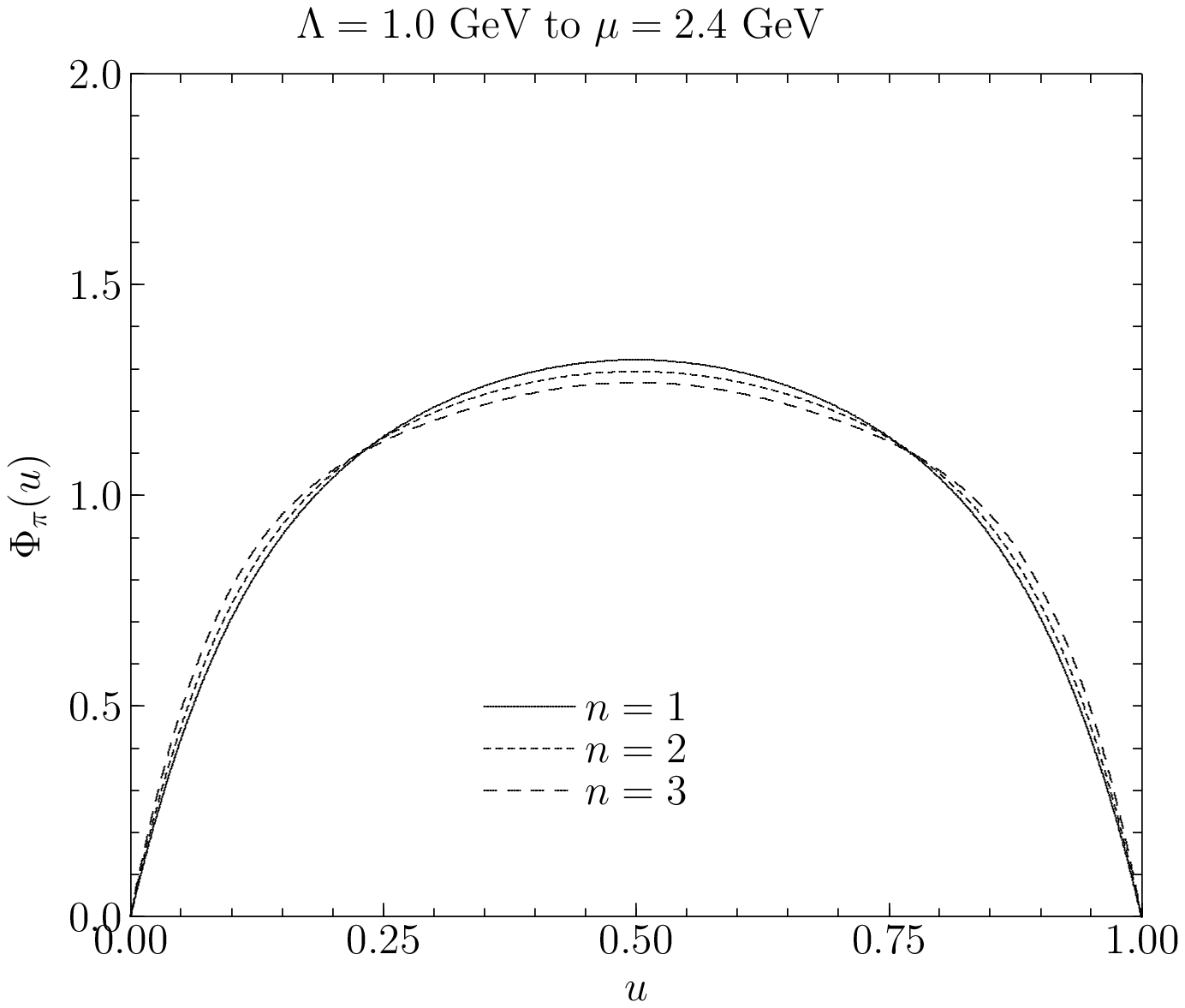} 
\includegraphics[width=5.5cm]{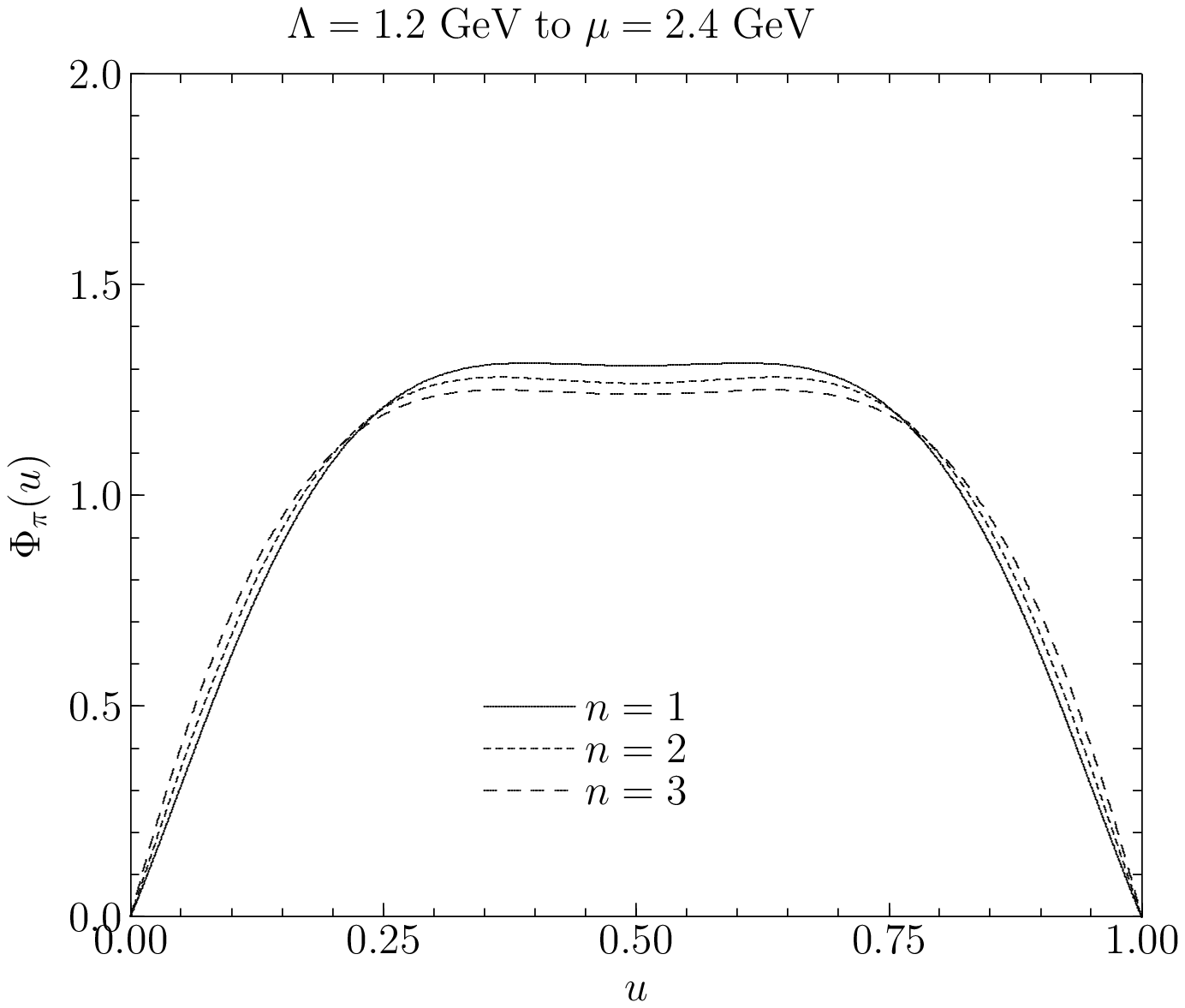} 
\includegraphics[width=5.5cm]{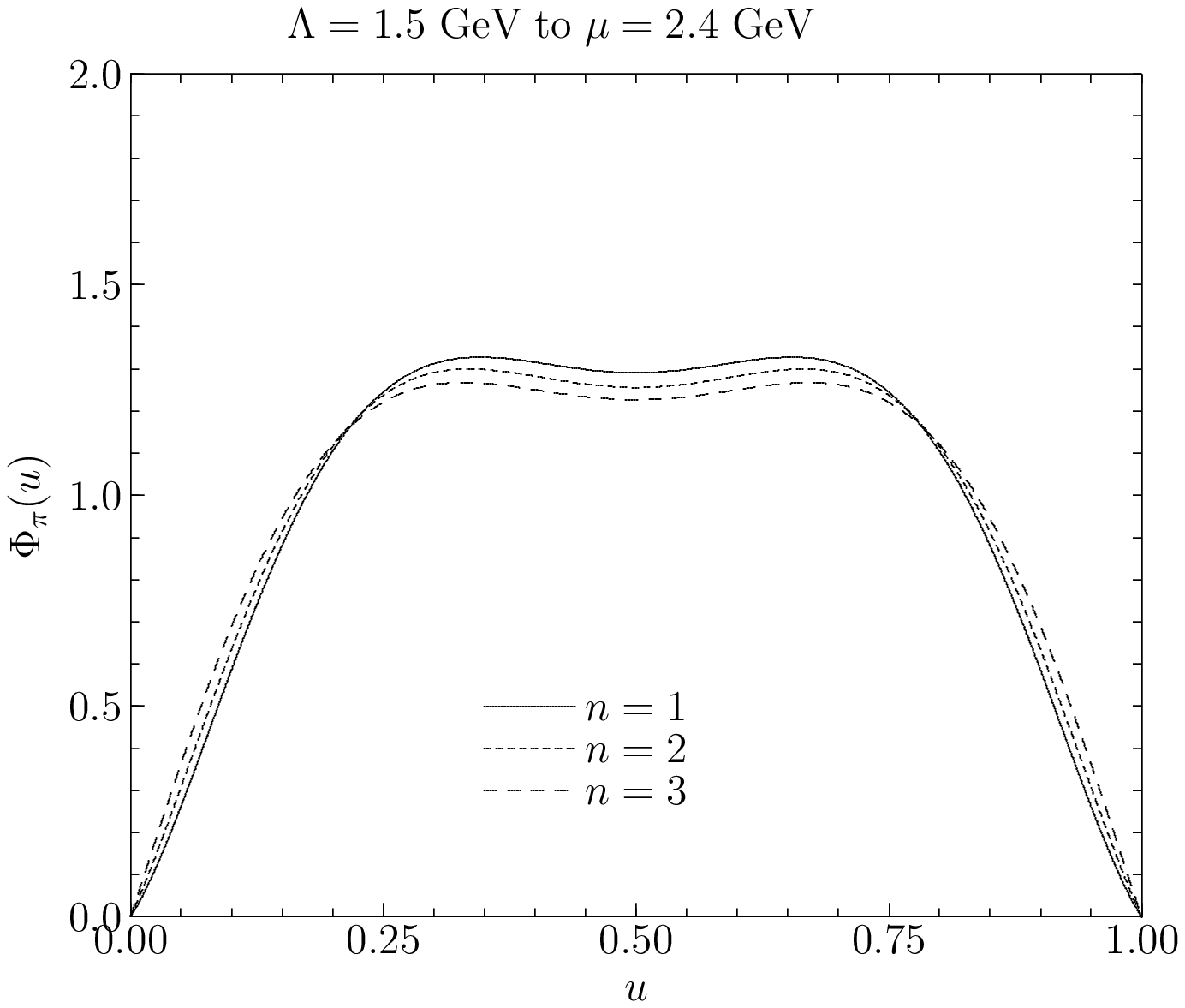} 
\end{tabular} 
\caption{Pion distribution amplitudes at $\mu=2.4$ GeV with $\Lambda=1.0$ GeV, 
  $1.2$ GeV and $1.5$ GeV from the left end panel.}  
\label{fig4} 
\end{figure}

To show the dependence of the pion DAs on the scale parameter 
$\Lambda$, we draw in Fig.~\ref{fig4} the pion DA with different 
$\Lambda$.  We observe that the pion DAs are more suppressed at the 
end points and more humped as $\Lambda$ increases.   
 
\section{Summary and Conclusion} 
We investigated the pion and kaon distribution amplitudes within the 
framework of the nonlocal chiral quark model from the instanton vacuum, 
making the gauge invariance of the low-energy partition function 
preserved in the presence of the external fields.  The scale parameter 
of the present work was set to be $\Lambda\simeq1.0$ GeV which is not
far from that determined by using the saddle-point equation.  Using
$\Lambda\simeq1.0$ GeV, we were able to reproduce the empirical values
of the pion and kaon decay constants qualitatively well by using the
normalization condition of the distribution amplitudes.  The effects
of flavor SU(3) symmetry breaking effects were explicitly taken into
account consistently.    
 
We found that the nonlocal contribution arising from the gauge 
invariance of the effective chiral action makes 
the pion distribution amplitude flattened more in the region of 
$0.25\lesssim{u}\lesssim0.75$, compared to the former investigation 
without consideration of the gauge invariance~\cite{Nam:2006au}.  The 
results indicate that the absolute values of Gegenbauer moments $a_2$ 
and $a_4$ become larger.  As for the kaon, the nonlocal contribution 
is less important.  We noticed that $a^K_1$ is still negative even 
with the nonlocal contributions, which is similar to those in the 
previous work~\cite{Nam:2006au}.   
  
By virtue of the renormalization group equation, we evolved the
Gegenbauer moments $a_2$ and $a_4$ so that we may compare them   
with the Schmedding-Yakovlev analysis on the CLEO experimental data at   
$\mu=2.4$ GeV.  We found that the present results with the gauge 
invariance turn out to satisfy the Schmedding-Yakovlev analysis of the 
Gegenbauer moments at the $95\%$ confidential level.  We found that 
the values of $(a^K_2,a^K_4)$ were very close to those of  
the asymptotic distribution amplitude.     
 
In conclusion, the present gauge-invariant approach has several 
advantages, compared to the former works.  Firstly, it provides a more 
reasonable scale to the pion and kaon DAs, since we can use the 
normalization condition without any gauge-invariant problem.    
Secondly, the nonlocal corrections arising from the gauge invariance 
of the effective chiral action play a significant role in describing
the pion distribution amplitude.  Compared to the Schmedding-Yakovlev
analysis, the present results are phenomenologically better than those
of the former work.  In order to describe the pion and kaon
distribution amplitudes more consistently, we want to introduce the
meson-loop corrections ($1/N_c$ corrections) to them.  The
corresponding work is under progress.   
      
\section*{Acknowledgments} 
The work of S.N. is supported by the Brain Korea 21 (BK21) project 
in Center of Excellency for Developing Physics Researchers of Pusan 
National University, Korea. S.N. would like to thank J.~H.~Lee for 
fruitful discussions.  
\section*{Appendix} 
The first term in the square parenthesis in Eq.~(\ref{DA3}) reads 
\begin{eqnarray} 
&&{\rm  tr}_{\gamma}
\left[\frac{\sqrt{M_f(k)}}{D(k)}\rlap{/}{\hat{n}}\gamma_5\frac{ 
\sqrt{M_f(k-P)}}{D(k-P)}\gamma_5\right]=-4M_0\sqrt{f(m_f)f(m_f)} 
\left[\frac{n\Lambda^2}{n\Lambda^2-k^2}\right]^{n}\left[\frac{ 
n\Lambda^2}{n\Lambda^2-(k-P)^2}\right]^{n}\nonumber\\&\times& 
\left\{k\cdot\hat{n}\left[m_g+M_0f(m_g)\left[\frac{n\Lambda^2}{ 
n\Lambda^2-(k-P)^2}\right]^{2n}\right]-(k-P)\cdot\hat{n}\left[ 
m_f+M_0f(m_f)\left[\frac{n\Lambda^2}{n\Lambda^2-k^2}\right]^{2n} 
\right]\right\}\nonumber\\&\times& 
\left\{k^2-m^2_f-2M_0m_ff(m_f)\left[\frac{n\Lambda^2}{n\Lambda^2-k^2} 
\right]^{2n}-M^2_0f^2\left[\frac{n\Lambda^2}{n\Lambda^2-k^2}\right]^{4n} 
\right\}^{-1}\nonumber\\&\times&  
\left\{(k-P)^2-m^2_g-2M_0m_gf(m_g)\left[\frac{n\Lambda^2}{n\Lambda^2-(k-P)^2} 
\right]^{2n}-M^2_0f^2\left[\frac{n\Lambda^2}{n\Lambda^2-(k-P)^2}\right]^{4n} 
\right\}^{-1}\nonumber\\&=&4\sqrt{\eta_f\eta_g}(\alpha 
k_--\gamma_f)^n(\beta k_--\gamma_g)^n\nonumber\\&\times&\frac{\alpha(\alpha 
  k_--\gamma_f)^{2n}[m_g(\beta k_--\gamma_g)^{2n}+\eta_g]-\beta (\beta 
  k_--\gamma_g)^{2n}[m_f(\alpha k_--\gamma_f)^{2n}+\eta_f]}{\mathcal{D}_f 
\mathcal{D}_g},\nonumber\\\nonumber 
\end{eqnarray} 
where 
\begin{eqnarray} 
\mathcal{D}_f&=&(\alpha k_--D_f)(\alpha k_--\gamma_f)^{4n}-2\eta_fm_f(\alpha k_--\gamma_f)^{2n}-\eta^2_f+i\epsilon,\nonumber\\  
\mathcal{D}_g&=&(Bk_--D_g)(Bk_--\gamma_g)^{4n}-2\eta_gm_g(Bk_--\gamma_g)^{2n}-\eta^2_g+i\epsilon.\nonumber 
\end{eqnarray} 
We use the following parameterizations for convenience: 
\begin{eqnarray} 
&&k^2=k_+k_--k^2_T=uP_+k_--k^2_T=\alpha k_--k^2_T,\nonumber\\ 
&&(k-P)^2=(u-1)P_+k_--(u-1)m^2_{\phi}-k^2_T=\beta 
k_--(u-1)m^2_{\phi}-k^2_T,\nonumber\\  
&&n\Lambda^2-k^2=-uP_+k_-+[k^2_T+n\Lambda^2]=-\alpha k_-+\gamma_f,\nonumber\\ 
&&n\Lambda^2-(k-P)^2=-(u-1)P_+k_-+[(u-1)m^2_{\phi}+k^2_T+n\Lambda^2]=-\beta 
k_-+\gamma_g,\nonumber\\ 
&&k^2-m^2_f=\alpha k_--[k^2_T+m^2_f]=\alpha k_--\delta_f,\nonumber\\ 
&&(k-P)^2-m^2_g=\beta 
k_--[(u-1)m^2_{\phi}+k^2_T+m^2_g]=\beta k_--\delta_g,\nonumber\\ 
&&M_0f(m_f)(n\Lambda^2)^{2n}=\eta_f,\,\,\,\,M_0f(m_g)(n\Lambda^2)^{2n}=\eta_g,\nonumber\\  
&&M_f(k)=\frac{\eta_f}{(\alpha k_--\gamma_f)^{2n}},\,\,\,\,M_f(k-P)=\frac{\eta_g}{(\beta k_--\gamma_g)^{2n}},\nonumber\\ 
&&\frac{1}{D(k)}=\frac{(\alpha 
  k_--\gamma_f)^{4n}}{\mathcal{D}_f}\left[\rlap{/}{k}+m_f+\frac{\eta_f}{(\alpha k_--\gamma_f)^{2n}}\right],\nonumber\\&&\frac{1}{D(k-P)}=\frac{(\beta 
  k_--\gamma_g)^{4n}}{\mathcal{D}_g}\left[(\rlap{/}{k}-\rlap{/}{P})+m_g+\frac{\eta_f}{(\beta k_--\gamma_g)^{2n}}\right].\nonumber 
\end{eqnarray} 
The second term can be evaluated 
using the dynamical quark mass of Eq.~(\ref{dynamicalmass}) as 
follows:  
\begin{eqnarray} 
&&2n{\rm tr}_{\gamma}\left[\frac{\hat{n}\cdot k\sqrt{M_f(k)}\sqrt{M_f(k-P)}}{D(k)(n\Lambda^2-k^2)}\right]=\frac{8n 
  k\cdot\hat{n}M_0\sqrt{f(m_f)f(m_g)}}{(n\Lambda^2-k^2)}\left[\frac{n\Lambda^2}{n\Lambda^2-k^2}\right]^{n}\left[\frac{n\Lambda^2}{n\Lambda^2-(k-P)^2}\right]^{n}\nonumber\\&\times&\left\{m_f+M_0f(m_f)\left[\frac{n\Lambda^2}{n\Lambda^2-k^2}\right]^{2n}\right\}\left\{k^2-m^2_f-2M_0f(m_f)\left[\frac{n\Lambda^2}{n\Lambda^2-k^2}\right]^{2n}-M^2_0f^2\left[\frac{n\Lambda^2}{n\Lambda^2-k^2}\right]^{4n}\right\}^{-1}\nonumber\\&=&- 
\frac{8n\alpha\sqrt{\eta_f\eta_g}\left[m_f(\alpha 
    k_--\gamma_f)^{2n}+E_f\right](\alpha k_--\gamma_f)^{n-1}}{(\beta k_--\gamma_g)^n\mathcal{D}_f},\nonumber  
\label{second} 
\end{eqnarray} 
Similarly, the third term can be written as follows: 
\begin{eqnarray}                         
&&2n{\rm tr}_{\gamma}\left[\frac{ 
    (k-P)\cdot\hat{n}\sqrt{M_f(k)}\sqrt{M_f(k-P)}}{D(k-P)[n\Lambda^2-(k-P)^2]}\right]= 
-\frac{8n\beta\sqrt{\eta_f\eta_g}\left[m_g(\beta 
    k_--\gamma_g)^{2n}+\eta_g\right](\beta k_--\gamma_g)^{n-1}}{(\alpha 
    k_--\gamma_f)^n\mathcal{D}_g}.\nonumber                   
\end{eqnarray}           
For convenience, we denote these three terms evaluated above by 
$\mathcal{T}^V_L$ and $\mathcal{T}^V_{NL}$, which is the sum of the second and 
third terms.  
                             
\end{document}